\documentclass[conference]{IEEEtran}
\IEEEoverridecommandlockouts

\normalsize
\usepackage{amsmath,amssymb,amsthm}
\usepackage{enumerate}
\usepackage{stfloats}
\usepackage{comment}
\usepackage{subcaption}
\usepackage{siunitx}
\usepackage{mathtools}
\usepackage{accents,latexsym,cancel}
\usepackage{cite}
\usepackage{stackengine}

\usepackage[dvipsnames]{xcolor}
\definecolor{myPink}{RGB}{255,105,183}

\usepackage[T1]{fontenc}
\usepackage{graphics} 
\usepackage{epsfig} 
\usepackage[mathscr]{euscript}
\usepackage{algorithm}
\floatstyle{plaintop}
\restylefloat{algorithm}
\usepackage[noend]{algpseudocode}
\usepackage{bbm}
\makeatletter
\def\BState{\State\hskip-\ALG@thistlm}
\makeatother

\usepackage{tikz}
\usetikzlibrary{arrows,shapes,chains,matrix,positioning,scopes,patterns,calc,
decorations.markings,
decorations.pathmorphing,
}

\usepackage{pgfplots}
\pgfplotsset{compat=1.3}
\usepgflibrary{shapes}

\renewcommand{\epsilon}{\varepsilon}

\newcommand{\RNum}[1]{\uppercase\expandafter{\romannumeral #1\relax}}

\newcommand{\mv}{\ensuremath{\mathbf{m}}}

\newcommand{\vv}{\ensuremath{\mathbf{v}}}

\newcommand{\wv}{\ensuremath{\mathbf{w}}}

\newcommand{\Am}{\ensuremath{\mathbf{A}}}
\newcommand{\Dm}{\ensuremath{\mathbf{D}}}

\newcommand{\Ktot}{\ensuremath{K_{\mathrm{tot}}}}
\newcommand{\Ka}{\ensuremath{K_{\mathrm{a}}}}

\def\Pr{\mathrm{Pr}}

\DeclareMathAlphabet{\mcl}{OMS}{cmsy}{m}{n}

\newlength\tikzwidth
\newlength\tikzheight

\definecolor{mycolor1}{rgb}{0.63529,0.07843,0.18431}%
\definecolor{mycolor2}{rgb}{0.00000,0.44706,0.74118}%
\definecolor{mycolor3}{rgb}{0.00000,0.49804,0.00000}%
\definecolor{mycolor4}{rgb}{0.87059,0.49020,0.00000}%
\definecolor{mycolor5}{rgb}{0.00000,0.44700,0.74100}%
\definecolor{mycolor6}{rgb}{0.74902,0.00000,0.74902}%

\usepackage{tikz}
\usepackage{fancyvrb}
\usepackage{graphicx}
\usepackage{algorithm}
\usepackage[noend]{algpseudocode}
\usepackage{amsmath}
\usetikzlibrary{arrows,shapes,chains,matrix,positioning,scopes,patterns,calc,decorations.pathmorphing,shadows}
\usepackage{pgfplots}
\tikzstyle{cnode}=[circle,minimum size=0.75cm,draw]
\tikzstyle{cgnode}=[circle,draw]
\tikzstyle{crnode}=[circle,draw]
\tikzstyle{conode}=[rectangle,draw,fill=black!15]
\tikzstyle{cpnode}=[circle,draw]
\tikzstyle{pnode}=[circle,draw,fill=black,inner sep=1.5pt]
\tikzstyle{rnode}=[rectangle,draw,outer sep=0pt]
\tikzstyle{fnode}=[rectangle,draw,minimum height=0.6cm,minimum width=1.1cm,fill=red!75,text centered]
\tikzstyle{prnode}=[rectangle,minimum height=0.6cm,minimum width=1.1cm,text centered,draw]
\tikzstyle{block}=[rectangle,draw,minimum width=2cm,minimum height=1cm,text centered]
\tikzstyle{sblock}=[rectangle,draw,minimum width=1cm,minimum height=1cm,text centered]
\tikzstyle{prnodesimple}=[rectangle,draw,text width=11em,text centered,outer sep=0pt]
\tikzstyle{bigsnake}=[snake=snake,segment amplitude=4mm, segment length=4mm, line after snake=5mm]
\tikzstyle{smallsnake}=[snake=snake,segment amplitude=0.7mm, segment length=4mm, line after snake=3mm]
\newcommand{\norm}[1]{\left\lVert#1\right\rVert}
\title{LDPC Codes with Soft Interference Cancellation for Uncoordinated Unsourced Multiple Access}
\author{\IEEEauthorblockN{
Asit Kumar Pradhan, \emph{Member, IEEE},
Vamsi K. Amalladinne, \emph{Student Member, IEEE}, \\
Krishna R. Narayanan, \emph{Fellow, IEEE},
Jean-Francois Chamberland, \emph{Senior Member, IEEE} \\
Department of Electrical and Computer Engineering, Texas A\&M University}
\thanks{
This material is based upon work supported, in part, by the National Science Foundation (NSF) under Grants CCF-1619085 and by Qualcomm Technologies, Inc., through their University Relations Program.
}
}

\begin{document}

\maketitle

\begin{abstract}
This article presents a novel enhancement to the random spreading based coding scheme developed by Pradhan et al.\ for the unsourced multiple access channel.
The original coding scheme features a polar outer code in conjunction with a successive cancellation list decoder (SCLD) and a hard-input soft-output MMSE estimator.
In contrast, the proposed scheme employs a soft-input soft-output MMSE estimator for multi-user detection.
This is accomplished by replacing the SCLD based polar code with an LDPC code amenable to belief propagation decoding.
This novel framework is leveraged to successfully pass pertinent soft information between the MMSE estimator and the outer code.
LDPC codes are carefully designed using density evolution techniques to match the iterative process.
This enhanced architecture exhibits significant performance improvements and represents the state-of-the-art over a wide range of system parameters.
\end{abstract}

\begin{IEEEkeywords}
Unsourced multiple access, spreading sequences, low-density parity-check codes, belief propogation
\end{IEEEkeywords}

\section{Introduction}

Next generation wireless networks are envisioned to support a massive number of sporadically communicating, unattended devices.
The traffic generated by such devices differs significantly from the datastreams produced by smartphones and similar mobile platforms.
Indeed, existing devices are typically attached to a human operator and, as such, the traffic they generate is characterized by sustained connections with long payload lengths.
The advent of machine-type communications (MTC) forms a considerable departure from this standing.
The unsourced multiple access (UMAC) paradigm~\cite{polyanskiy2017perspective} 
models the transmission of a massive number of infrequently transmitting devices.

In the conventional MAC setting, the base station is cognizant of active users and each user employs a different codebook to transmit information.
However, in the UMAC model, the total device population in a network is huge, orders of magnitude larger than the number of active devices.
This reality, together with short payloads, renders traditional decoding impractical because the identities of the active devices must be resolved as part of the communication process.
The UMAC paradigm circumvents this issue by allowing active devices to share a common codebook.
User identity plays no role in the encoding process and decoding is performed only up to a permutation of the transmitted message sequences.
If needed, active devices can embed their own identity as a part of the payload.
These aspects are discussed in great detail and a random coding achievability bound is derived for the UMAC channel in the absence of complexity constraints in~\cite{polyanskiy2017perspective}.

Since the introduction of this benchmark, there have been considerable research efforts in designing low-complexity solutions that offer good performance tailored to the UMAC setting~\cite{ordentlich2017low,vem2019user,amalladinne2019coded,Giuseppe,amalladinne20,pradhan2019joint,pradhan2019polar,calderbank2018chirrup,marshakov2019polar,zheng2020polar,amalladinne2019enhanced,amalladinne2019asynchronous}.
These algorithms can be broadly classified into two categories: compressed sensing (CS) based approaches and coding-based approaches.
Within the former approaches, the UMAC is viewed as the support recovery of a sparse vector whose length grows exponentially with the payload.
Standard CS solvers cannot be directly applied to this problem because their complexity grows linearly with vector length.
In~\cite{amalladinne2019coded,Giuseppe,amalladinne20}, this issue is addressed by first splitting the payloads into smaller pieces, performing CS recovery on the sub-problems, and subsequently stitching pieces together using an outer code.
Next, we discuss the main ideas in the coding based schemes, which rely on sparsifying collisions to reduce multi-user interference (MUI).
In~\cite{ordentlich14,vem2019user,marshakov2019polar}, schemes in the spirit of $T$-fold ALOHA are proposed.
Specifically, the approach therein is to divide the transmission period into slots and to limit the expected number of users that transmit within each slot.
In~\cite{pradhan2019joint}, an interleave division multiple access (IDMA) based scheme is considered, which aims at reducing MUI at the bit level.
This scheme achieves the aforementioned objective by first encoding the payload using a code whose blocklength is much smaller than the transmission period, and then interleaving the resulting codeword into the whole frame.

Another coding-based approach leverages spreading sequences to control multiple access interference.
Such a scheme using polar codes and random spreading was proposed in \cite{pradhan2019polar}.
In this scheme, the payload corresponding to each active user is split into two parts.
The first component acts as a preamble to choose a signature sequence from a dictionary of sequences with good correlation properties.
The remaining bits are encoded using a polar code whose frozen bits are dictated by the preamble.
Decoding begins with an energy detector that leverages the correlation properties of these sequences to identify the subset of selected spreading sequences.
The preambles corresponding to active users are implicitly decoded in this step.
This information is passed to a minimum mean squared error (MMSE) estimator that produces log-likelihood ratios (LLRs) for the symbols appearing in the residual part of the messages based on each active spreading sequence.
These LLRs act as input to a single-user polar list decoder that attempts to recover the latter part of the messages, treating interference as noise.
Finally, the successfully decoded polar codewords are removed from the received signal in the spirit of successive interference cancellation, and the residual signal is fed back to the energy detector for the next iteration.
This scheme is state-of-the-art when the number of active devices is less than $225$. 
Yet, its performance deteriorates drastically as the number of users increases; it underperforms when the active population exceeds $250$ users.

In this article, we propose an enhancement to the scheme in \cite{pradhan2019polar} which shows significant performance improvements, thereby making this new variant the best performing practical scheme to date.
Notably, the proposed scheme outperforms the random coding achievability bound derived in \cite{polyanskiy2017perspective} when the active population is less than $75$.
Further, the proposed scheme substantially outperforms the scheme in \cite{pradhan2019polar}
when the active population exceeds $200$ users.
We list below the key features that distinguish our proposed scheme from previous implementations.
\begin{itemize}
    \item In \cite{pradhan2019polar}, a polar outer code is used in conjunction with a successive cancellation list decoder (SCLD). 
    The SCLD does not naturally produce soft output and, hence, a hard-input soft-output MMSE estimator was used for the multi-user detector.
    In this paper, we replace the hard-input MMSE estimator with a soft-input soft-output (SISO) MMSE estimator. 
    To enable passing soft information between the estimator and the outer decoder, we replace the polar code with an (low-density parity-check) LDPC outer code that is decoded using belief propagation.
    The use of a SISO estimator leverages the availability of information from partially decoded codewords and improves the decoding process.
    This produces significant overall performance gains.
         \item The polar codes used in~\cite{pradhan2019polar}, while powerful, are not designed to integrate an iterative decoding and estimation process.
     This explains the poor performance of the scheme when the number of active users is large. 
     However, in our proposed scheme, LDPC codes are carefully crafted using density evolution to be matched to the iterative process.
     The performance gains of our proposed scheme come at the expense of increased complexity.
     Specifically, its decoding complexity is an order higher than the one in~\cite{pradhan2019polar}.
 \end{itemize}
Throughout, we adopt the following notation.
The sets $\mathbb{R}$, $\mathbb{Z}$ represent the real numbers and the integers, respectively. 
We denote the set of integers from $i$ to $j$, inclusively, by $i : j$.
We use boldface capital letters for matrices and underlined variables for vectors. 
We denote the $(i,j)$th element of matrix $\mathbf{A}$ by $A_{i,j}$.
The $i$th row and the $j$th column of matrix $\mathbf{A}$ are represented by $\mathbf{A}_{i,:}$ and $\mathbf{A}_{:,j}$, respectively.
We refer to the $i$th element of vector $\underbar v$ by $v(i)$.
For vectors $\underbar u, \underbar v \in \mathbb{R}^n$, the operator $\langle\underbar u, \underbar v\rangle$ denotes the standard inner product.

\section{System Model}
Consider a wireless network with a total of $\Ktot$ devices among which a small subset $\Ka$ are active at a given time.
Each active device has a payload of $B$ bits that it intends to convey to a central base station.
A transmission interval of $n$ real channel uses is allocated towards this end.
The signal received by the central base station takes the form
\begin{equation}
\label{eq:1}
\underbar y = \sum_{k=1}^{\Ka} \underbar x_{k}(\underbar{w}_{k}) + \underbar z,
\end{equation}
where $\underbar{w}_{k} \in \{0,1\}^B$ denotes the payload corresponding to active user $k$, and $\underbar x_{k}(\underbar{w}_{k}) \in \mathbb{R}^n$ is the signal transmitted by active user $k$.
Noise vector $\underbar z$ has independent components, each with Gaussian distribution $\mathcal{N} (0, \sigma^2)$. 
The signals transmitted by active users are subject to individual power constraint $\|\underbar x_{k}(\underbar{w}_{k})\|^2 \le n$.
The operating energy per bit to noise power spectral density ratio is defined as $\frac{E_b}{N_0}=\frac{n}{2B\sigma^2}$.
The base station is tasked with recovering the collection of transmitted messages given the received signal $\underbar y$.
Towards this end, the receiver constructs a list of decoded messages $\mathcal{L}(\underbar y)$ whose size does not exceed $\Ka$, i.e., $|\mathcal{L}(\underbar y)| \le \Ka$. 
The design objective is minimizing the SNR subject to a target per-user error probability $P_e$, which is defined as
\begin{equation}\label{eqn:proboferrordefinition}
  P_e = \frac{1}{\Ka} \sum_{k=1}^{\Ka}\Pr\left( \underbar w_k \notin \mathcal{L}(\underbar y) \right).
\end{equation}
It is worth mentioning that this has become the predominant performance criterion for UMAC schemes.

\section{Proposed Scheme}
\label{Sec:ProposedScheme}
\begin{figure}
    \centering
    \begin{tikzpicture}[
  font=\footnotesize, >=stealth', line width = 0.75pt,
  block/.style={rectangle, minimum height=8mm, minimum width=12mm, draw=black}
]

\node[block] (message) at (0,0) {$\underbar w=(\underbar w_{\mathrm{s}},\underbar w_{\mathrm{c}})$};
\draw[->] (-1.25,0) to (message);

\node[block] (A) at (3.75,0) {$\mathbf{A}$}; 
\draw[->] (message) to node[above]{$\underbar w_{\mathrm{s}} \in \{0,1\}^{B_{\mathrm{p}}}$} (A);

\node[block] (chcode) at (0,-2) {LDPC}; 
\draw[->] (message) to node[right]{$\underbar w_{\mathrm{c}} \in \{0,1\}^{B_{\mathrm{c}}}$} (chcode);

\node[block] (bpsk) at (3.15,-2) {BPSK};
\draw[->] (chcode) to node[below]{$\underbar u \in \{0,1\}^{n_\mathrm{c}}$} (bpsk);

\node[block] (spreading) at (5.75,-1) {$\underbar v \otimes \underbar a_j$};
\draw[->] (A) to node[above]{$\underbar a_j \in \mathbb{R}^{n_{\mathrm{p}}}$} (5.75,0) to (spreading);
\draw[->] (bpsk) to node[below]{$\underbar v \in \{-1,1\}^{n_\mathrm{c}}$} (5.75,-2) to (spreading);
\draw[->] (spreading) to node[above] {$\underbar x$} (6.75,-1);

\end{tikzpicture} 
    \caption{This figure illustrates the encoding process.
    A message is partitioned into two parts. The first part selects a signature sequence from a dictionary for spreading. The second part is encoded using a powerful error correcting code.}
    \label{fig:encoder}
\end{figure}
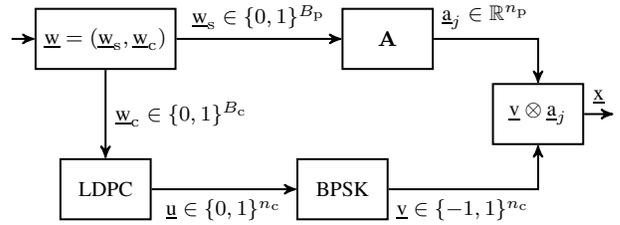

We present a detailed description of the proposed framework in this section.
The broad structure of the encoder used in this scheme remains the same as the one developed in \cite{pradhan2019polar}.
Still, for the sake of completeness, we briefly outline the encoding process, which is also encapsulated through a notional diagram in Fig.~\ref{fig:encoder}.
We then describe the various components used within the iterative decoding process.

\subsection{Encoder}
\label{sec:encoder}
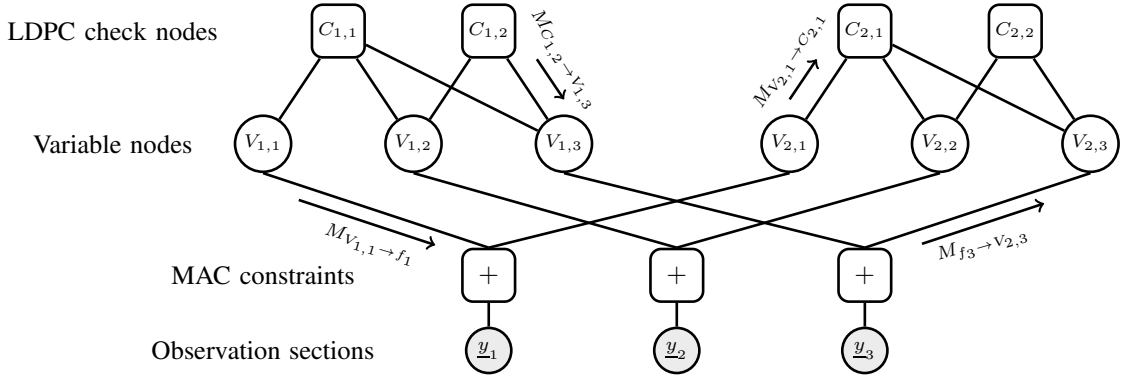
\begin{figure*}[tbh]
    \centering
    \begin{tikzpicture}[font=\scriptsize, line width = 1pt,
    cnode/.style={rectangle, minimum width=7mm, minimum height=7mm, draw=black, inner sep=2pt,rounded corners},
    fnode/.style={rectangle, minimum width=4.5mm, minimum height=4.5mm, draw=black},
    vnode/.style={circle, minimum size=7mm, draw=black, inner sep=2pt},
    ynode/.style={circle, minimum size=6mm, draw=black, fill=lightgray!30, inner sep=2pt},
]

\foreach \j in {1,2} {
    \foreach \i in {1,2} {
        \node[cnode] (C\j\i) at (2*\j + 7*\i - 8,2) {$C_{\i,\j}$};
    }
}
\node (check) at (-2,2) {\normalsize LDPC check nodes};

\foreach \j in {1,2,3} {
    \foreach \i in {1,2} {
        \node[vnode] (V\j\i) at (2*\j + 7*\i - 9,0.5) {$V_{\i,\j}$};
    }
}
\node (variable) at (-2,0.5) {\normalsize Variable nodes};

\foreach \j in {1,2,3} {
    \foreach \i in {1,2} {
        \draw (V\j\i) to (C1\i);
    }
}
\foreach \j in {2,3} {
    \foreach \i in {1,2} {
        \draw (V\j\i) to (C2\i);
    }
}

\foreach \s in {1,2,3} {
    \node[cnode] (m\s) at (2.5*\s + 0.5, -1.25) {\normalsize $+$};
    \node[ynode] (y\s) at (2.5*\s + 0.5, -2.25) {$\underline{y}_{\s}$}
        edge (m\s);
}
\node (mcheck) at (0,-1.25){\normalsize MAC constraints};
\node (observations) at (0,-2.25){\normalsize Observation sections};

\foreach \s in {1,2,3} {
    \foreach \i in {1,2} {
        \draw (V\s\i.south) to (m\s.north);
    }
}

\draw[shorten <=0.8cm,shorten >=0.65cm,->] ([yshift=-0.2cm]m3.north) -- node[below,rotate=16] {\scriptsize $M_{f_3 \rightarrow V_{2,3}}$} ([yshift=-0.2cm]V32.south);

\draw[shorten <=0.5cm,shorten >=0.75cm,->] ([yshift=-0.2cm]V11.south) -- node[below,rotate=-16] {\scriptsize $M_{V_{1,1} \rightarrow f_1}$} ([yshift=-0.2cm]m1.north);

\draw[shorten >=0.25cm,->] ([yshift=0.2cm]V12.north) -- node[above,rotate=57.5] {\scriptsize $M_{V_{2,1} \rightarrow C_{2,1}}$} ([xshift=-0.5cm,yshift=0.2cm]C12.south);

\draw[shorten <=0.25cm,->] ([xshift=0.5cm,yshift=0.2cm]C21.south) -- node[above,rotate=-57.5] {\scriptsize $M_{C_{1,2} \rightarrow V_{1,3}}$} ([yshift=0.2cm]V31.north);


\end{tikzpicture}
    \caption{This figure shows a Tanner graph induced by two users.
    Each received symbol is represented by a MAC node.
    Since there are two users, every MAC constraint is connected to two variable nodes.
    }
    \label{fig: decoder }
\end{figure*}

The $B$-bit message corresponding to each active user is split into two parts of length $B_{\mathrm{p}}$ and $B_{\mathrm{c}}=B-B_{\mathrm{p}}$, respectively.
We use the term preamble to refer to the first portion of the message, $\underbar{w}_{\mathrm{p}}$; the remaining bits are denoted by $\underbar{w}_{\mathrm{c}}$.
The $B_{\mathrm{p}}$-bit preamble is employed to select a column from matrix $\mathbf{A}=[\underbar{a}_1, \underbar{a}_2,\ldots,\underbar{a}_{2^{B_{\mathrm{p}}}}] \in \mathbb{R}^{n_{\mathrm{p}} \times 2^{B_{\mathrm{p}}}}$, which then acts as a signature sequence for this particular user.
The choice function is bijective, a property which ensures that the preamble can also be recovered from the signature sequence.
Message component $\underbar{w}_{\mathrm{c}}$ is encoded using an LDPC code and the resulting codeword, denoted by $\underbar u$, is modulated using BPSK to produce a vector $\underbar{v} \in \{-1, 1\}^{n_{\mathrm{c}}}$.
Constant $n_{\mathrm{p}}$ and codelength $n_{\mathrm{c}}$ are such that $n_{\mathrm{c}} = \lfloor n/n_{\mathrm{p}} \rfloor$.
For convenience, we denote the spreading sequence picked by the $k$th active user by $\underbar{a}_{j_k}$.
Let $\mathcal{B}$ represent the set of indices chosen by at least one active user, i.e., $j' \in \mathcal{B}$ if $\exists k$ such that $j_k = j'$.
Finally, the modulated codeword is spread using the chosen signature to produce signal transmit $\underbar x$.
For active user $k$, this signal assumes the form
\begin{equation} \label{eq:3}
    \underbar {x}_k= \underbar v_k \otimes \underbar{a}_{j_k},
\end{equation}
where $\otimes$ denotes the tensor product operation.
We note that this architecture induces a Tanner graph at the receiver, as illustrated in Fig.~\ref{fig: decoder }.
At this point, we are ready to describe the decoding process.

\subsection{Decoder}
\label{sec:decoder}
The iterative decoder features several components.
Its operation is summarized as follows.
First, an energy detector is used to identify the signature sequences, or equivalently preambles, of the transmitted messages.
The inactive columns are then pruned from $\mathbf{A}$.
Based on the compact form, a SISO MMSE estimator is employed to produce estimates of the codeword symbols associated with each active spreading sequence.
The symbol estimates corresponding to each active sequence are then passed to a single-user decoder.
This decoder leverages the structure of the code to improve the symbol estimates corresponding to every sequence returned by the energy detector.
With these refined estimates, the decoder loops back to the SISO MMSE estimator, setting the stage for an iterative process.

Since the decoding process is impervious to user identities, we describe the algorithmic structure in terms of spreading sequences.
We begin by giving a matrix representation of the received signal.
Recall that multiple users can pick the same spreading sequence if they share a common preamble. 
We denote the collection of active users who select sequence $\underbar a_j$ by $\mathcal{M}_j$.
We represent the sum of all codewords associated with sequence $\underbar a_j$ by $\tilde{\underbar{v}}_j$, with
\begin{equation}
    \label{eq:vtilde}
    \tilde{\underbar{v}}_j \coloneqq \sum_{k \in \mathcal{M}_j}\underbar v_k
\end{equation}
and $\tilde{\underbar{v}}_j$ is the zero vector when $\mathcal{M}_j = \emptyset$.
This collection of vectors, $\tilde{\underbar{v}}_j$ for $1 \leq j \leq 2^{B_{\mathrm{p}}}$, can be arranged in the form of a matrix,
\begin{equation*}
\widetilde{\mathbf{V}}
= \begin{bmatrix}
\tilde{\underbar{v}}_1^{\mathsf{T}} \\ \tilde{\underbar{v}}_2^{\mathsf{T}} \\
\vdots \\ \tilde{\underbar{v}}_{2^{B_{\mathrm{p}}}}^{\mathsf{T}}
\end{bmatrix}.
\end{equation*}
As we will see shortly, the juxtaposition of column vectors $\tilde{\underbar{v}}_j$ and rows in $\widetilde{\mathbf{V}}$ (which may be surprising) is a byproduct of spreading.
We introduce a symbol partitioning for $\underbar y$; the section of the received signal corresponding to the $i$th LDPC symbol is
\begin{equation*}
    \underbar {y}_i= \begin{bmatrix} \underbar y((i-1)n_{\mathrm{p}}+1) & \underbar y((i-1)n_{\mathrm{p}}+2) & \cdots & \underbar y(in_{\mathrm{p}})\end{bmatrix}^{\mathsf{T}}.
\end{equation*}
The received vector $\underbar y$ can be reshaped into a matrix $\mathbf{Y}$,
$$\mathbf{Y}= \begin{bmatrix} \underbar{y}_1 & \underbar{y}_2 & \cdots & \underbar{y}_{n_{\mathrm{c}}}\end{bmatrix}.$$ 
We can partition $\underbar z$ in the same manner, leading to section $\underbar{z}_i$ being the noise vector affecting $\underbar{y}_i$.
With this extended notation, $\underbar y_i$ in \eqref{eq:1} can be expressed as
\begin{xalignat}{3} \label{equation:SectionObseravtions}
\underbar y_i &= \mathbf{A} \widetilde{\mathbf{V}}_{:,i} + \underbar{z}_i, i=1,\ldots,n_{\mathrm{c}} &
&\text{or} & \mathbf{Y} &= \mathbf{A} \widetilde{\mathbf{V}} + \mathbf{Z} .
\end{xalignat}
We emphasize that, in matrix notation, $\mathbf{Y}_{:,i} = \underbar{y}_i$ is the section of the observation linked to the $i$th LDPC symbol in each codeword.
These observed values are affected by all the users, hence the presence of $\widetilde{\mathbf{V}}_{:,i}$ in \eqref{equation:SectionObseravtions}.

The operations involved in each decoding iteration are similar; we therefore explain the decoding for one iteration.
Belief propagation for our decoder involves several types of messages.
We introduce the following convenient notation that delineates the source and destination of the messages.
\begin{center}
\renewcommand{\arraystretch}{1.1}
\begin{tabular}{l}
$M_{{V}_{j,i} \rightarrow C_{j,l}}$: from variable node ${V}_{j,i}$ to check node $C_{j,l}$ \tabularnewline
$M_{C_{j,l} \rightarrow {V}_{j,i}}$: from check node $C_{j,l}$ to variable node ${V}_{j,i}$ \tabularnewline
$M_{{V}_{j,i} \rightarrow f_{i}}$: from variable node ${V}_{j,i}$ to factor node $f_{i}$ \tabularnewline
$M_{f_i \rightarrow {V}_{j,i}}$: from factor node $f_i$ to variable node ${V}_{j,i}$
\end{tabular}
\end{center}
It should be kept in mind that index $j$ refers to the $j$th spreading sequence and index $i$ refers to the $i$th coded bit.
We are ready to describe the various components of the decoder in detail.

\subsubsection{Active Spreading Sequence Detector (SSD)}
The first step in the receiver is to form an estimate of the indices of the active users or, equivalently, their spreading sequences.
 We refer the reader to \cite{pradhan2019polar} for a detailed description of this detector and we only summarize the main idea below.
For each sequence $\underbar a_j$, we employ a matched-filter receiver and produce a decision statistic given by 
\begin{equation*}
   \nu_j = \norm{ \underbar a_j^{\mathsf{T}}\mathbf{Y}}^2.
\end{equation*}
The list of $\nu_j$s is sorted in descending order and the
SSD outputs the top $\Ka$ sequences from the ordered list.
 We denote the set  of sequence indices returned by the SSD by $\mathcal{D}$.
 While it is desirable for $\mathcal{D}$ to be identical to the actual set of indices corresponding to the active users $\mathcal{B}$, in a practical receiver, the two sets may differ.
Note that detecting the active sequences is equivalent to decoding the preamble part of the messages.
\subsubsection{SISO MMSE Estimator}
In this step, the MAC nodes take the LLRs corresponding to the codeword symbols from the LDPC decoder as soft input.
Then, it performs SoIC and MMSE estimation to refine the estimates of the codeword symbols using the algorithm of Wang and Poor~\cite{wang1999iterative}.
A brief description of the algorithm follows.

Let $\mathbf{S} = \mathbf{A}_{:,\mathcal{D}}$ denote the matrix $\mathbf{A}$ restricted to the columns indexed by elements in the set $\mathcal{D}$ and let
$\widetilde{\mathbf{V}}_{\mathcal{D,:}}$ denote the matrix $\widetilde{\mathbf{V}}$ restricted to row indices from $\mathcal{D}$.
We approximate the matrix $\mathbf{Y}_{:,i}$ by
\begin{equation}
\label{eq:4}
    \mathbf{Y}_{:,i} \approx \mathbf{S}\widetilde{\mathbf{V}}_{\mathcal{D},i}+\underbar{z}_i,
\end{equation}
where 
The entries of vector $\underbar z_i$ are independent zero-mean Gaussian random variable with variance $\sigma^2$.
In practical settings, undetected sequences produce interference, which is implicitly treated as noise.
Thus, \eqref{eq:4} should be interpreted as a good approximation in the presence of undetected sequences, and it becomes precise when $\mathcal{D}$ is recovered exactly.
It can be seen from \eqref{eq:vtilde} that if spreading sequence $\mathbf{a}_j$ is chosen by only active user, then $\tilde{\underbar{v}}_j = \underbar{v}_j$. 
We describe the message passing rules only for such spreading sequences in order to keep the description clear. 
Therefore, messages are labeled by $V_{j,i}$ instead of $\widetilde{V}_{j,i}$.
Nevertheless, we implement the same message passing algorithm for spreading sequences chosen by multiple users too.

A MAC node first forms  soft estimates of the codeword symbols based on the extrinsic information obtained from the single user channel decoder $M_{V_{j,i} \rightarrow f_i}$.
For instance, the $i$th symbol corresponding to the $j$th sequence is estimated by the conditional mean estimate, which for BPSK modulation becomes
\begin{equation}
\label{eq:5}
   \widehat{{V}}_{j,i} = \tanh\left(\frac{M_{{V}_{j,i} \rightarrow f_{i}}}{2}\right).
\end{equation}
Initially, the messages from variable nodes to factor nodes are set to $M_{V_{j,i} \rightarrow f_i}=0$.
The message passing from variable nodes to factor nodes is discussed in the next section.
These soft estimates are used to perform SoIC to obtain effective observations corresponding to the codeword symbols.
After the SoIC, the effective observation of the codeword symbol $V_{j,i}$ can be written as
\begin{equation}
\begin{split}
\label{eq:6}
    { \mathbf{Y}}_{:,i}^j &=  \mathbf{Y}_{:,i} -  \sum_{l\in \mathcal{D},l \neq j} \mathbf{S}_{:,l}\widehat{{V}}_{l,i}, \\ 
    &= \mathbf{S}_{:,j}{{V}}_{j,i} + \sum_{l \in \mathcal{D},l\neq j} \mathbf{S}_{:,l}\left({V}_{l,i} - \widehat{{V}}_{l,i}\right) + \underbar{z}_i.
\end{split}
\end{equation}
Let $\mathbf{P}^{j,i}$ be a diagonal matrix whose $(l,l)$th entry is
$\mathbb{E}[({V}_{l,i}-\hat{V}_{l,i})^2]$ given by
\begin{equation}
\mathbf{P}_{l,l}^{j,i}=
\begin{cases}
  \label{eq:7}
      1 -\left(\tanh \left(\frac{M_{V_{l,i} \rightarrow f_{i}}}{2}\right)\right)^2 & \text{ for } l \neq j \\
     1 & \text{ for } l = j .
\end{cases}
\end{equation}
With that, \eqref{eq:6} can be approximated by
\begin{align}
    {\mathbf{Y}}_{:,i}^j \approx \mathbf{S}\sqrt{\mathbf{P}^{j,i}} {\mathbf{V}}_{\mathcal{D},i} +\underbar z_i,
\end{align}
We pass ${\mathbf{Y}}_{:,i}^j$ through a linear minimum mean-squared error (MMSE) filter to obtain an estimate of ${{V}}_{j,i}$ given by
\begin{align}
\label{eq:9}
    \widehat{\mathbf{T}}_{j,i}= \mathbf{S}_{:,j}^{\mathsf{T}}\left(\mathbf{S}\mathbf{P}^{j,i}\mathbf{S}^{\mathsf{T}}+\sigma^2\mathbf{I}\right)^{-1}{\mathbf{Y}}_{:,i}^{j} .
\end{align}
The mean squared error (MSE) of this estimator is given by
\begin{align}
\label{eq:10}
    \gamma^2_{j,i}= 1 - \mathbf{S}_{:,j}^{\mathsf{T}}\left(\mathbf{S}\mathbf{P}^{j,i}\mathbf{S}^{\mathsf{T}}+\sigma^2\mathbf{I}\right)^{-1}\mathbf{S}_{:,j}.
\end{align}
Observe that $\widehat{\mathbf{V}}_{j,i}$'s are the soft inputs to the multi-user MMSE detector and $\widehat{\mathbf{T}}_{j,i}$'s are the soft outputs from it.
To proceed with the iterative decoding process, we need to compute $M_{{f_i} \rightarrow {V}_{j,i}}$ from $\widehat{{V}}_{j,i}$, where $M_{f_{i} \rightarrow V_{j,i}}$ is the LLR of the codeword symbol ${V}_{j,i}$.
In this context, $\widehat{{V}}_{j,i}$ can be seen as the MMSE estimate of ${V}_{j,i}$ when it is passed through an equivalent  AWGN channel with noise variance $\sigma^2_{\mathrm{eq}}$. 
Mathematically 
\begin{align*}
    y_{\mathrm{eq}}= {{V}}_{j,i} + z_{\mathrm{eq}},
\end{align*}
where $z \sim \mathcal{N}(0, \sigma^2_{\mathrm{eq}})$, $\sigma^2_{\mathrm{eq}}=\frac{\gamma^2_{j,i}}{1-\gamma_{j,i}^2}$, and $y_{\mathrm{eq}}=\frac{\widehat{{T}}_{j,i}}{1-\gamma_{j,i}^2}$.
Then, the LLRs at the output of the equivalent AWGN channel can be computed as
\begin{align}
\label{eq:11}
M_{f_{i} \rightarrow {V}_{j,i}} = \frac{2\widehat{{T}}_{j,i}}{\gamma^2_{j,i}},
\end{align}
and forwarded to the LDPC decoder.

As described in \eqref{eq:9}, this step requires inversion of a $n_{\mathrm{p}} \times n_{\mathrm{p}}$ matrix to produce the soft estimate of a codeword symbol, whose complexity is $\mathcal{O}(n_{\mathrm{p}}^3)$.
 Since there are $\Ka$ users and each of them has $n_{\mathrm{c}}$ symbols, the complexity of this step is $\mathcal{O}(\Ka n_{\mathrm{c}} n_{\mathrm{p}}^3 )$. 
 The proposed scheme performs well when $n_\mathrm{p}$ is $\mathcal{O}(\Ka)$. 
 Henceforth, the complexity of this step can be approximated to $\mathcal{O}(n_{\mathrm{c}} \Ka^4)$. In \cite{pradhan2019polar}, the multi-user MMSE estimator is the same for all the coded symbols and active sequences. 
 Therefore, it requires the inversion of a single matrix. So the complexity of this step is $\mathcal{O}(\Ka^3).$ 
 Here, we also remark that the complexity of the other decoding steps is linear in both $\Ka$ and $n_{\mathrm{c}}$.    
\subsubsection{Channel Decoder}

The message passing rules at the variable nodes and check nodes of the LDPC code are standard and well known \cite{richardson2008modern}. For sake of completeness, we review them here.
Let $\mathcal{N}_{V_{j,i}}$ and $\mathcal{N}_{C_{j,l}}$ denote the set of nodes connected to 
variable node $V_{j,i}$ and check node $C_{j,l}$, respectively. 
At the $i$th varibale node of the $j$th user, we have 
\begin{align}
\label{eq:12}
    M_{{V}_{j,i} \rightarrow C_{j,l}}&= \sum_{C \in \mathcal{N}_{{V}_{j,i}} \setminus C_{j,l}} M_{C \rightarrow {V}_{j,i}} + M_{f_{i} \rightarrow {V}_{j,i}},\\
    \label{eq:13}
    M_{{V}_{j,i} \rightarrow f_{i}} &= \sum_{C \in \mathcal{N}_{{V}_{j,i}} } M_{C \rightarrow {V}_{j,i}}.
\end{align}
The message passing rule at the $l$th check node of the $j$th user is given by
\begin{align}
\label{eq:14}
    M_{C_{j,l} \rightarrow {V}_{j,i}} = 2 \tanh^{-1}\left( \prod_{{V} \in \mathcal{N}_{C_j}\setminus {V}_{j,i} } \tanh (M_{{V} \rightarrow C_{j,l}}) \right).
\end{align}
At the end of a fixed number of iterations, an estimate of  LDPC codeword corresponding to the $j$th sequence is obtained by 
\begin{align*}
    \hat{\underbar{u}}_j(i) = \begin{cases}
    1, \text{ if } \sum_{C \in \mathcal{N}_{{V}_{j,i}}} M_{C \rightarrow {V}_{j,i}} + M_{f_{i} \rightarrow {V}_{j,i}} >0\\
    0, \text{ otherwise}.
    \end{cases}
\end{align*}
Let $\overline {\mathcal{D}}$ denote the set of indices such that $\hat{\underbar{u}}_j$ is a valid LDPC codeword, i.e., all the parity checks are satisifed for each $j \in \overline {\mathcal{D}}$.
This is the set of indices that are deemed to be successfully decoded by the LDPC decoder. 
The SIC removes the contributions from all the valid LDPC codewords $\mathbf{V}_{\widetilde{\mathcal{D}}}$ from the received signal to compute the residual
\begin{equation}
\label{eq:15}
{\mathbf{Y}}_{i,:} - \sum_{j \in \overline{\mathcal{D}}} \mathbf{S}_{:,j}(1-2\hat{u}_j(i)),
\end{equation}
 for $1 \leq i \leq n_{\mathrm{c}}.$
This residual is passed back to the energy detector for the second iteration.
This process continues until all the transmitted messages are recovered successfully or there is no improvement between two consecutive rounds of the iterative process.
We encapsulate the overall decoding process in Algorithm~\ref{alg:decoding}.

\begin{algorithm}[t]
\begin{algorithmic}[1]
\State Energy detector returns an estimate of the set of active sequences $\mathcal{D}$.
\State Initialize $M_{V_{j,i} \rightarrow f_i}=0$, $\forall j$ and $\forall i$. 
\State Compute $\widehat{\mathbf{V}}_{j,i}$ using \eqref{eq:5} $\forall j$ and $\forall i.$
\State Perform SoIC at the MAC nodes using \eqref{eq:6}.
\State Compute $\widehat{\mathbf{T}}_{j,i}=\mathbf{S}_{:,j}^{\mathsf{T}}\left(\mathbf{S}\mathbf{P}\mathbf{S}^{\mathsf{T}}+\sigma^2\mathbf{I}\right)^{-1}\widetilde{\underbar y}_{j,i}.$
\State Compute MSE $\gamma^2_{j,i}= 1 - \mathbf{S}_{:,j}^{\mathsf{T}}\left(\mathbf{S}\mathbf{P}\mathbf{S}^{\mathsf{T}}+\sigma^2\mathbf{I}\right)^{-1}\mathbf{S}_{:,j}.
$
\State Convert $\widehat{\mathbf{T}}_{j,i}$ into LLR using \eqref{eq:11}.
\State Compute $M_{V_{j,i} \rightarrow C_{j,l}}$ using \eqref{eq:12}.
\State Compute $M_{C_{j,l} \rightarrow V_{j,i}}$ using \eqref{eq:14}.
\State Compute $M_{V_{j,i} \rightarrow f_{i}}$ using \eqref{eq:13}.
\State Repeat steps 2 -- 10 for a fixed number of times.
\State Update $\overline{\mathcal{D}}$ as the set of  indices corresponding to valid LDPC codewords.
\State Perform SIC using \eqref{eq:15}.
\State Repeat steps 1--13 until all the active users are decoded or  $\overline{\mathcal{D}} = \varnothing$ 
\end{algorithmic}
\caption{\rule[-.3\baselineskip]{0pt}{1.35\baselineskip} Decoding Algorithm}\label{alg:decoding}
\end{algorithm}

\section{Simulation Results}
\label{section:simulationResults}
\begin{figure}
\centering
  \begin{tikzpicture}
\definecolor{mycolor1}{rgb}{0.63529,0.07843,0.18431}%
\definecolor{mycolor2}{rgb}{0.00000,0.44706,0.74118}%
\definecolor{mycolor3}{rgb}{0.00000,0.49804,0.00000}%
\definecolor{mycolor4}{rgb}{0.87059,0.49020,0.00000}%
\definecolor{mycolor5}{rgb}{0.00000,0.44700,0.74100}%
\definecolor{mycolor6}{rgb}{0.74902,0.00000,0.74902}%
\definecolor{mycolor7}{rgb}{0.502,0.2000,0.5902}

\begin{axis}[%
font=\footnotesize,
width=7.5cm,
height=7cm,
scale only axis,
xmin=25,
xmax=275,
xtick = {25,75,...,275},
xlabel={\small Number of active users $\Ka$},
xmajorgrids,
ymin=0,
ymax=6,
ytick = {0,2,...,6},
ylabel={\small Required $E_b/N_0$ (dB)},
ylabel near ticks,
ymajorgrids,
legend style={font=\footnotesize, at={(0.86,1)},anchor=north east, draw=black,fill=white,legend cell align=left}
]

\addplot [color=black,dotted,line width=1.5pt]
  table[row sep=crcr]{
 25	0.25\\
50	0.3\\
75	0.35\\
100	0.4\\
125	0.45\\
150	0.5\\
175	0.55\\
200	0.6\\
225	0.95\\
250	1.25\\
275	1.55\\
300	1.8\\
};
\addlegendentry{Random Coding (Polyanskiy '17)};

\addplot [color=mycolor3,solid,line width=1.5pt,mark size=1.0pt,mark=square,mark options={solid}]
  table[row sep=crcr]{
  25  2\\
50	2.1\\
75	2.2\\
100	2.41\\
125	2.57\\
150	2.81\\
175	3\\
200 3.4\\
225 3.88\\
250 4.36\\
275 4.87\\
300 5.35\\
};\addlegendentry{Sparse IDMA, (Pradhan, et al. '19)};

\addplot [color=mycolor6,solid,mark=*,line width=2.0pt]
  table[row sep=crcr]{
  2     0.3\\
  10    0.5\\
 25	0.55\\
50	0.6\\
75 0.7\\
100 0.75\\
125 1.15\\
150 1.5\\
175 2\\
200 2.7\\
225 3.5\\
250 4.3\\
};
\addlegendentry{Polar+Spreading (Pradhan, et al. '20)};
\addplot [color=mycolor7,solid,line width=2.0pt,mark size=1.4pt,mark=o,mark options={solid}]
  table[row sep=crcr]{10 1.75\\
  25  2.05\\
50	2.15\\
75	2.37\\
100	2.47\\
125	2.75\\
150	3.25\\
175 3.6\\
200 3.8\\
225 3.95\\
250 4.4\\
275 4.65\\
300 5.1\\
};
\addlegendentry{Enhanced AMP+Tree (Amalladinne et al. '20)};
\addplot [color=mycolor4,solid,line width=2.0pt,mark size=1.4pt,mark=triangle,mark options={solid}]
  table[row sep=crcr]{
  25  0.73\\
50	1.15\\
75	1.63\\
100	2.06\\
125	2.48\\
150	2.99\\
175	3.43\\
200 3.89\\
225 4.41\\
250 4.91\\
275 5.26\\
300 5.49\\
};
\addlegendentry{IRSA + Polar Coding (Marshakov et al. '20)}

\addplot [color=cyan,solid,line width=2.0pt,mark size=1.4pt,mark=o,mark options={solid}]
  table[row sep=crcr]{
  25  0.1\\
  50 0.1\\
  75 0.3
  100 0.4\\
  125 0.55\\
  150 1\\
  175 1.3\\
200 1.65\\
225 2.35\\
250 2.85\\
275 3.43\\
};
\addlegendentry{Spreading+soft-cancellation};

\end{axis}
\end{tikzpicture}%
  \caption{The figure compares the performance of the proposed scheme with existing schemes. The proposed scheme outperforms the state-of-the-art.}
  \label{fig:results}
\end{figure}
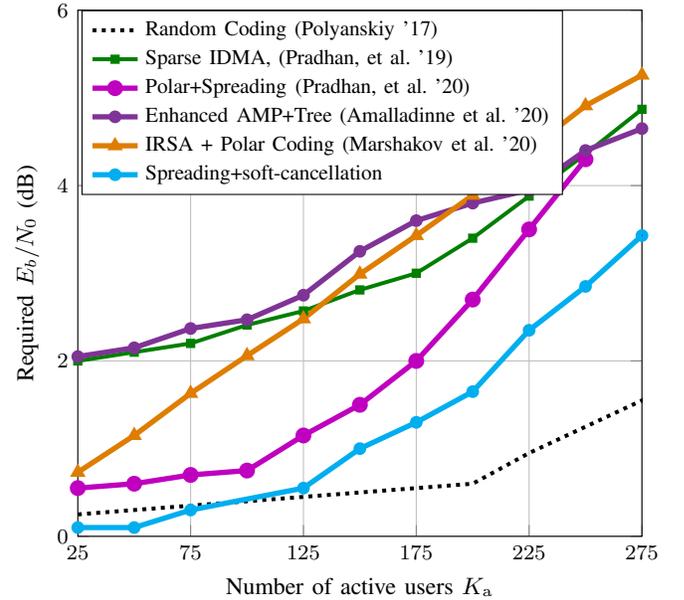

We compare the minimum $E_b/N_0$ required to achieve a target per user probability of error $P_e=0.05$ for the proposed scheme to that of existing schemes that have been designed for UMAC. 
For this purpose, we choose the total number of channel uses  $n=30000$, and the message size $B=100$ bits. 
The parameters used in simulations are given below.
\begin{itemize}
    \item We choose $B_{\mathrm{p}}=12$ and $B_{\mathrm{c}}=B-B_{\mathrm{p}}=88$. Note that the signature sequence matrix $\mathbf{A}$ has $2^{B_{\mathrm{p}}}$ columns.
    \item We choose the length of the spreading sequence $n_{\mathrm{p}}=84.$
    \item The rate of the LDPC code is given by $R=\frac{B-B_{\mathrm{p}}}{\lfloor n|n_{\mathrm{p}}\rfloor}=0.25.$
\end{itemize}
These parameters are optimized empirically for $\Ka=250$ to minimize the energy-per-bit required to achieve a target probability of error.
Since the users pick sequences independently from $\mathbf{A}$, multiple users can choose the same sequence. We refer to such events as collisions. 
When collisions occur, we decode the stronger user by treating the interference due to other users as noise.
The LDPC codes are optimized using density evolution as described in \cite{Liva}.
Figure~\ref{fig:results} shows the performance comparison between the proposed schemes and previously published methods in the literature. 
The obtained simulation results show that the proposed scheme outperforms existing approaches over a wide range of $\Ka$. 
For $\Ka \leq 75$, the simulated performance is slightly better than the finite-blocklength (FBL) achievability bound developed in \cite{polyanskiy2017perspective}.
This is the first time performance better than the FBL achievability bound has been demonstrated.
For $75 \leq \Ka < 125$, 
simulated performance is comparable to that of the FBL achievability. 
For $125 < \Ka < 250$, the simulated performance is within 2~dB of the FBL achievability.


\bibliographystyle{IEEEbib}
\bibliography{IEEEabrv,MAC_collision}

\end{document}